\documentclass[letter,bibyear]{aa} % for the letters

\usepackage{epsfig}
\usepackage{amsmath}
\usepackage{subfigure}
\usepackage{natbib}
\usepackage{multirow}
\usepackage{color}
\usepackage{longtable}
\usepackage{graphicx}
\usepackage{epstopdf}
\usepackage{booktabs}
\usepackage{txfonts}
\newcommand{\jmst}{J.~Mol.~Struct.}

\newcommand{\diez}{10$^{10}$\,cm$^{-2}$}
\newcommand{\once}{10$^{11}$\,cm$^{-2}$}
\newcommand{\doce}{10$^{12}$\,cm$^{-2}$}

\begin{document}

\title{Discovery of two new interstellar molecules with QUIJOTE$^1$: HCCCHCCC and HCCCCS\thanks{Based on 
observations carried out
with the Yebes 40m telescope (projects 19A003,
20A014, 20D023, 21A011, and 21D005). The 40m
radio telescope at Yebes Observatory is operated by the Spanish Geographic Institute
(IGN, Ministerio de Transportes, Movilidad y Agenda Urbana).}}

\author{
R.~Fuentetaja\inst{1},
M.~Ag\'undez\inst{1},
C.~Cabezas\inst{1},
B.~Tercero\inst{2,3},
N.~Marcelino\inst{2,3},
J.~R.~Pardo\inst{1},
P.~de~Vicente\inst{2},
J.~Cernicharo\inst{1}
}

\institute{Dept. de Astrof\'isica Molecular, Instituto de F\'isica Fundamental (IFF-CSIC),
C/ Serrano 121, 28006 Madrid, Spain. \newline \email r.fuentetaja@csic.es, jose.cernicharo@csic.es
\and Centro de Desarrollos Tecnol\'ogicos, Observatorio de Yebes (IGN), 19141 Yebes, Guadalajara, Spain.
\and Observatorio Astron\'omico Nacional (OAN, IGN), C/ Alfonso XII, 3, 28014, Madrid, Spain.
}

\date{Received; accepted}

\abstract{We report on the discovery of two new molecules, HCCCHCCC and HCCCCS, towards 
the starless core TMC-1 in the Taurus region from the QUIJOTE line survey in the 31.1-50.2
GHz frequency range. We identify a total of twenty-nine lines of HCCCHCCC and six rotational transitions of HCCCCS. 
The rotational 
quantum numbers range from $J_u$=10 up to 15 and $K_a\leq$2 for HCCCHCCC
and $J_u$=21/2 up to 31/2 for HCCCCS. 
We derived a column density for HCCCHCCC of $N$=(1.3$\pm$0.2)$\times$10$^{11}$ cm$^{-2}$ with 
a rotational temperature of 6$\pm$1 K, while for HCCCCS we derived $N$=(9.5$\pm$0.8)$\times$10$^{10}$ 
cm$^{-2}$ and T$_{rot}$=10$\pm$1 K. The abundance of HCCCHCCC is higher than that of its recently discovered isomer, l-H$_2$C$_6$. If we compare HCCCCS with its related molecules, HCS and HCCS, we obtain abundance ratios HCS/HCCCCS=58 and HCCS/HCCCCS=7.2. We investigated the formation of these two molecules using chemical modelling calculations. The observed abundances can be accounted for by assuming standard gas-phase formation routes involving neutral-neutral reactions and ion-neutral reactions.
}


\titlerunning{HCCCHCCC and HCCCCS in TMC-1}
\authorrunning{Fuentetaja et al.}

\maketitle

\section{Introduction}
The number of molecules discovered in the cold dark cloud TMC-1, both with the Yebes 40m radio telescope through the 
QUIJOTE\footnote{\textbf{Q}-band \textbf{U}ltrasensitive \textbf{I}nspection \textbf{J}ourney 
to the \textbf{O}bscure \textbf{T}MC-1 \textbf{E}nvironment} line survey \citep{Cernicharo2021a} and the Green Bank 100m radio telescope with the GBT Observations of TMC-1: Hunting Aromatic Molecules (GOTHAM) survey 
\citep{McGuire2018}, demonstrates the great importance of this source for a complete understanding of the chemistry 
of the interstellar medium (ISM). Chemical models are not yet accurate enough to predict all the molecules that have 
been discovered, so we must continue to study the various chemical reactions in order to better constrain models of cold pre-stellar cores.

The TMC-1 cloud is remarkably rich in hydrocarbons, such as long carbon chains, propargyl \citep{Agundez2021}, vinyl acetylene \citep{Cernicharo2021b}, and allenyl diacetylene 
\citep{Fuentetaja2022}. Furthermore, several cyclic molecules, such as indene, cyclopentadiene \citep{Cernicharo2021c}, ortho-benzyne \citep{Cernicharo2021a}, and fulvenallene \citep{Cernicharo2022}, have been discovered there.

The carbenes, a family of molecules discovered in the ISM, are characterized by having 
two non-bonded 
electrons located on a terminal carbon atom. The most relevant are the cumulene carbenes, whose formula is 
H$_2$C$_n$. They have a chain with double bonds between their carbon atoms. The simplest and the first 
to be 
discovered was $l$-H$_2$C$_3$, in 1991 \citep{Cernicharo1991}. That same year, $l$-H$_2$C$_4$ was also discovered in TMC-1 
\citep{Kawaguchi1991} and IRC+10216 \citep{Cernicharo1991b} . The longest species detected so far is 
$l$-H$_2$C$_6$, which has been found in several sources, including TMC-1 \citep{Langer1997} and IRC+10216 \citep{Guelin2000}. 
Finally, \citet{Cabezas2021} detected $l$-H$_2$C$_5$ in TMC-1.
For all of these  molecules, the most stable isomer corresponds to the non-polar species with the formula 
HC$_n$H for an even n or a cyclic structure when n is odd. Several H$_2$C$_n$ isomers have been detected in 
the ISM, such as $c$-C$_3$H$_2$ \citep{Thaddeus1987} and $c$-C$_3$HCCH \citep{Cernicharo2021c}. 

In the case of the C$_6$H$_2$ family, \citet{sattelmeyer2000} performed the first theoretical study of the 
nine most stable isomers. They performed several types of ab initio calculations, the highest level being 
CCSD(T), using the cc-pVTZ basis set. The bent-chain isomer ethynylbutatrienyliden (HCCCHCCC) is 2.346 eV 
higher than the most stable isomer HC$_6$H, but only 0.1648 eV above the l-H$_2$C$_6$ 
species detected by \citet{Langer1997}. Moreover, it should be noted that there are two isomers whose 
energy is lower than those of these two species. The formula is c-C$_6$H$_2$ for both of them, with the hydrogens in the ortho 
and meta positions (1.7085 eV and 2.0078 eV above the more stable isomer, respectively), making them possible 
candidates for future discovery.

On the other hand, if we consider sulphur-bearing species, we see that the number of molecules 
detected in TMC-1 
is small compared with C- and O-rich species. However, some of them, such as CCS or CCCS, present a 
high abundance \citep{Saito1987,Yamamoto1987}. 
Among the last sulphur-bearing molecules discovered in TMC-1 we have HC$_3$S$^+$ \citep{Cernicharo2021d}, H$_2$CCS, 
H$_2$CCCS, NCS, HCCS, C$_4$S, and C$_5$S \citep{Cernicharo2021e}, the 
cation radical HCCS$^+$ \citep{Cabezas2022}, and the two derivatives of thioformaldehyde,
HCSCN and HCSCCH \citep{Cernicharo2021f}

In this Letter we report on the first identification of HCCCHCCC and HCCCCS towards TMC-1. We derive column densities 
and rotational temperatures for the two molecules and analyse them with our chemical models to explain which 
reactions can produce them.

\begin{figure}
\centering
\includegraphics[width=0.4\textwidth]{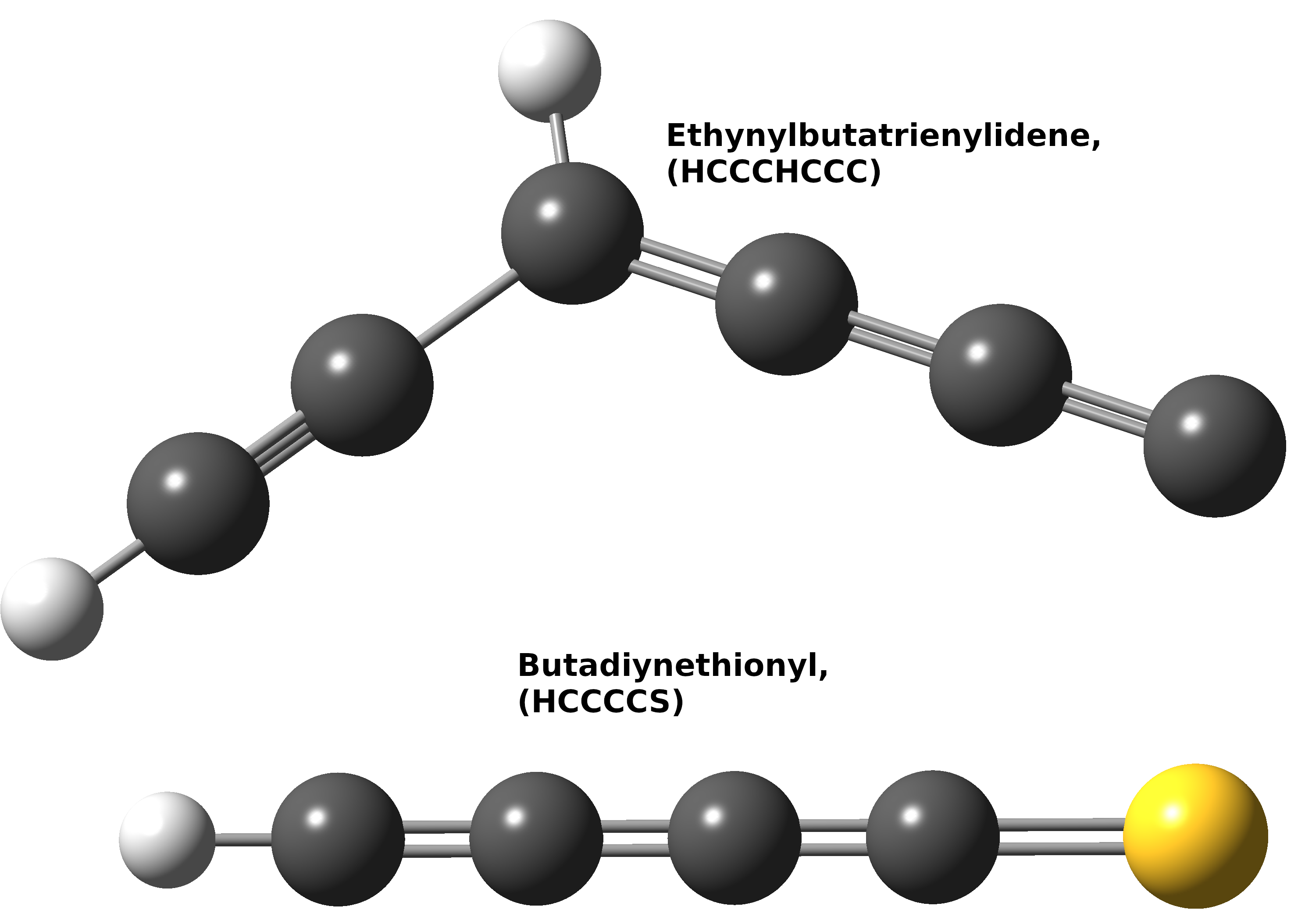}
\caption{Structure of the two molecules detected in this work.}
\label{fig_structur}
\end{figure}

\begin{figure*}
\centering
\includegraphics[width=0.9\textwidth]{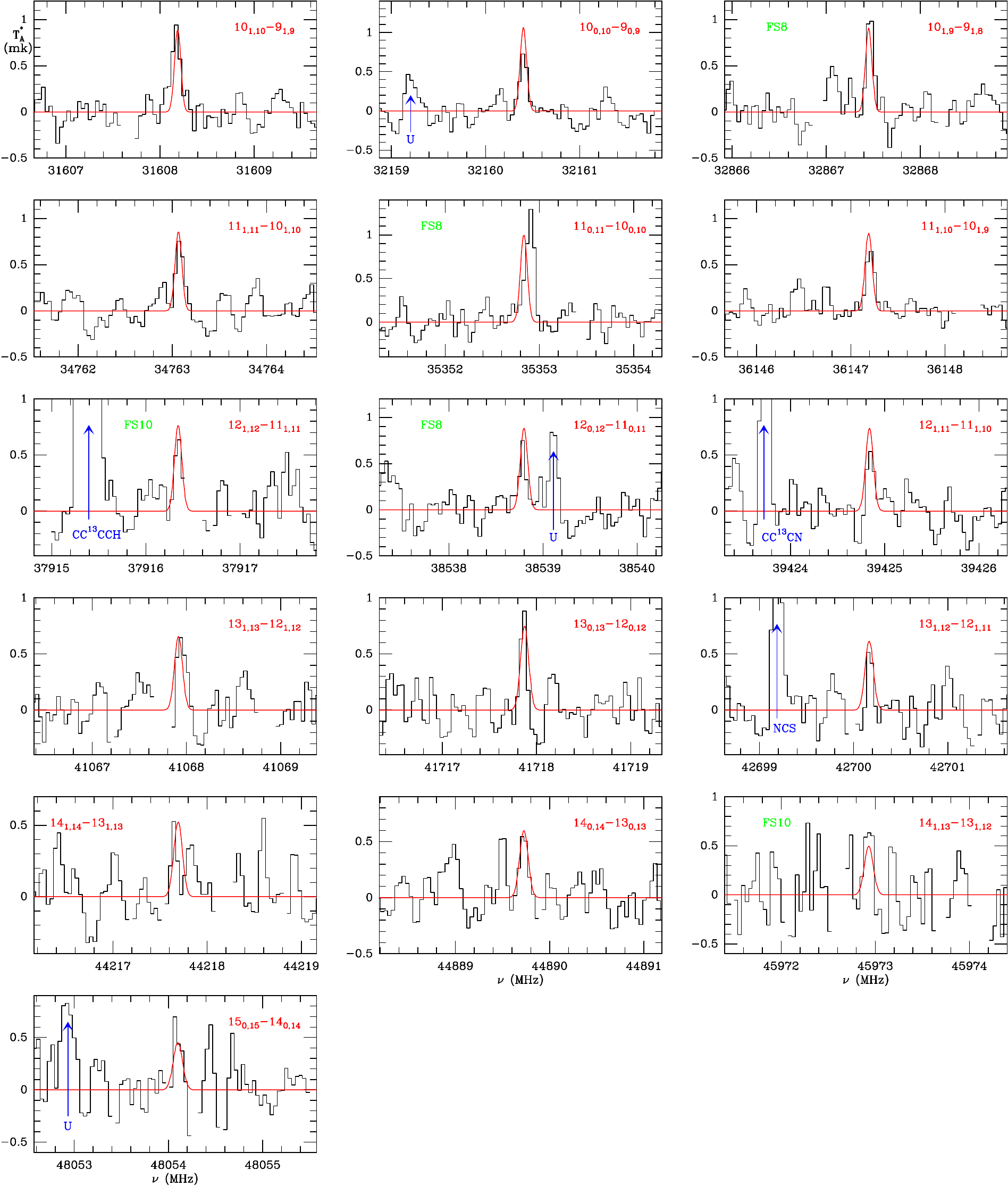}
\caption{Observed transitions $K_a$=0 and 1 of HCCCHCCC in TMC-1.
The abscissa corresponds to the rest frequency of the lines. Frequencies and intensities for the observed lines
are given in Table \ref{obs_line_parameters}.
The ordinate is the antenna temperature, corrected for atmospheric and telescope losses, in millikelvin.
The quantum numbers of each transition are indicated
in the corresponding panel. The red line shows the computed synthetic spectrum for this species for 
T$_{\mathrm{rot}}$ = 6$\pm$1 K and a column density of (1.3$\pm$0.2)$\times$\once. Blanked channels correspond to negative features 
produced when folding the frequency-switched data.  Green labels indicate the transitions for which 
only one of the frequency-switching data have been used (FS10 and FS8 correspond to a throw of 10 and 8 MHz, respectively).
}
\label{HCCCHCCC_k01}
\end{figure*}

\section{Observations}
\label{observations}

The observational data used in this work are part of the ongoing QUIJOTE line survey, carried out with the Yebes 40m telescope towards the cyanopolyyne peak of TMC-1 ($\alpha_{J2000}=4^{\rm h} 41^{\rm  m} 41.9^{\rm s}$ and $\delta_{J2000}=+25^\circ 41' 27.0''$). We used a new receiver built 
within the Nanocosmos project\footnote{\texttt{https://nanocosmos.iff.csic.es/}}, which consists of two cold high electron mobility transistor amplifiers covering the 31.0-50.3 GHz band 
with horizontal and vertical polarizations. Receiver temperatures achieved in the 2019 and 2020 runs
vary from 22 K at 32 GHz to 42 K at 50 GHz. Some power adaptation in the down-conversion chains 
effectively reduced the receiver temperatures in 2021 to 16\,K at 32 GHz and 30\,K at 50 GHz. 
The backends are $2\times8\times2.5$ GHz fast Fourier transform spectrometers with a spectral 
resolution of 38.15 kHz, providing full coverage of the Q band in both polarizations. A detailed 
description of the system is given by \citet{Tercero2021}.

The data come from several observing runs carried out between December 2019 and May 2022, and correspond to 546 
hours of observing time on source, of which 293 and 253 hours were acquired with a frequency-switching throw of 8 MHz and 10 MHz, respectively. A detailed description of the QUIJOTE line survey 
is provided in \citet{Cernicharo2021a}. For each frequency throw, different local oscillator frequencies 
were used in order to remove possible side band effects in the down-conversion chain.

The telescope 
beam size is 56$''$ and 31$''$ at 31 and 50 GHz, respectively.
The intensity scale used in this work, antenna temperature
($T_A^*$), was calibrated using two absorbers at different temperatures and the
atmospheric transmission model ATM \citep{Cernicharo1985, Pardo2001}.
The adopted calibration uncertainties are at a level of  10\%. The main beam efficiency varies from 0.6 at 32 GHz to 0.43 at 50 GHz \citep{Tercero2021}.
All the data were analysed using the GILDAS package\footnote{\texttt{http://www.iram.fr/IRAMFR/GILDAS}}.

\begin{figure*} 
\centering
\includegraphics[width=0.9\textwidth]{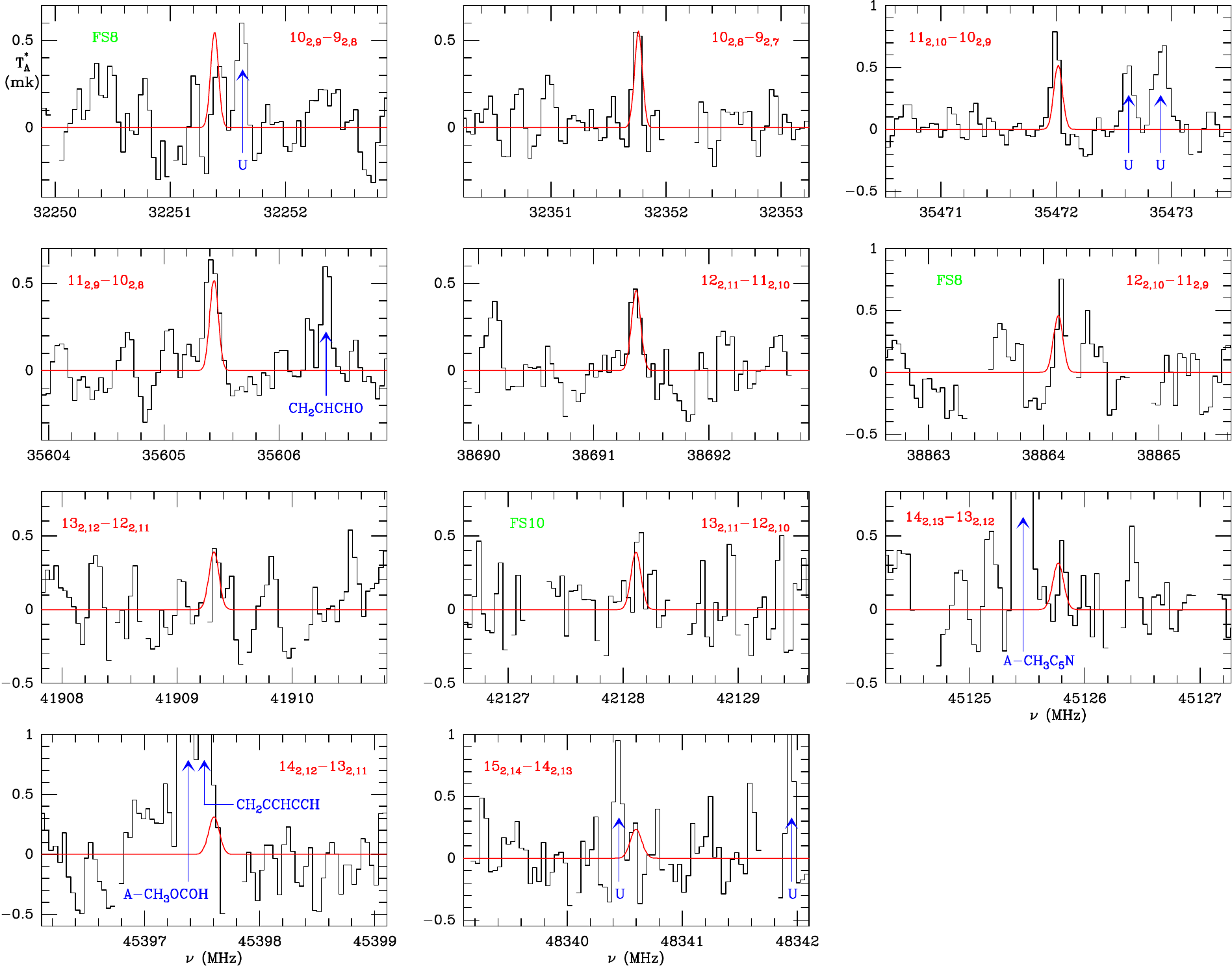}
\caption{Selected $K_a$=2  transitions of HCCCHCCC in TMC-1.
The abscissa corresponds to the rest frequency of the lines. Frequencies and intensities for all observed lines
are given in Table \ref{obs_line_parameters}.
The ordinate is the antenna temperature, corrected for atmospheric and telescope losses, in millikelvin.
The quantum numbers of each transition are indicated
in the corresponding panel. The red line shows the computed synthetic spectrum for this species 
for T$_{\mathrm{rot}}$ = 6$\pm$1 K and
a column density of (1.3$\pm$0.2)$\times$\once. Blanked channels correspond to negative features 
produced when folding the frequency-switched data. Green labels indicate the transitions for which 
only one of the frequency-switching data have been used (FS10 and FS8 correspond to a throw of 10 and 8 MHz, respectively).
}
\label{HCCCHCCCk2}
\end{figure*}

\section{Results}
\label{results}

For line identification, we used various catalogues, including CDMS \citep{Muller2005}, JPL \citep{Pickett1998}, 
and MADEX \citep{Cernicharo2012}, which, as of September 2022, contained a total of 6434 spectral entries 
corresponding to the isotopologues and the ground and excited states of 1734 molecules.

\subsection{Discovery of HCCCHCCC}

Ethynylbutatrienyliden is a planar molecule with a nearly prolate top symmetry and whose electronic ground state corresponds 
to a singlet $^1A$ state (see Fig. \ref{fig_structur}). It
was studied in the laboratory up to 40.3 GHz by \citet{McCarthy2002}, who determined with high precision its rotational 
and centrifugal distortion constants (see Table \ref{new_rotational_HCCCHCCC}). 
The dipole moments, $\mu_a$=3.710D and $\mu_b$=1.320D, were obtained by \citet{sattelmeyer2000}. 
We fitted these lines and implemented the rotational constants and the dipole moment into MADEX \citep{Cernicharo2012}. 
Twenty-nine lines of this species were found in our QUIJOTE data, with intensities between 0.5 and 1.2 mK.
Derived line parameters are given in Table \ref{obs_line_parameters} and the lines for $K_a$~$\leq$1 are shown
in Fig. 
\ref{HCCCHCCC_k01}, while those corresponding to $K_a$=2 are shown in Fig. \ref{HCCCHCCCk2}. 
Some lines are blended, but the large number of them allows for a solid detection.
From the observed frequencies in TMC-1, together with those measured by \citet{McCarthy2002}, we obtained a new 
set of rotational constants (see Appendix \ref{new_constants} and Table \ref{new_rotational_HCCCHCCC}) with the FITWAT code, which uses a Watson asymmetric rotor Hamiltonian in
$I^r$ representation \citep{Cernicharo2018}. The merged fit provides very good frequency predictions below 50 GHz and
realistic ones in the 50-100 GHz frequency range.

We used a model line fitting procedure \citep{Cernicharo2018} to derive the column density and the rotational 
temperature assuming all rotational levels have the same T$_{rot}$. We adopted a source of uniform brightness with 
a diameter of 80$''$ from the C$_6$H  maps presented in \citet{Fosse2001}.
We obtained T$_{rot}$=6$\pm$1\,K and $N$(HCCCHCCC)=(1.3$\pm$0.2)$\times$\once. To provide molecular abundances, we also assumed a H$_2$ column density for TMC-1 of 10$^{22}$cm$^{-2}$ \citep{Cernicharo1987}, so the abundance of HCCCHCCC is (1.3$\pm$0.2)$\times$10$^{-11}$.
Figures \ref{HCCCHCCC_k01} and \ref{HCCCHCCCk2} show the synthesized spectrum obtained with these values.
The column density for the ortho species $l$-H$_2$C$_6$ is 6$\times$\diez, which, together with
 2$\times$\diez\ for the para species, gives us a total column density $N$($l$-H$_2$C$_6$)=8$\times$\diez \citep{Cabezas2021}. Hence, the abundance
ratio HCCCHCCC/$l$-H$_2$C$_6$ is 1.625. This species, with an angle between the carbon atoms of 123.63º \citep{sattelmeyer2000}, could be a precursor of cyclic species.
   
\begin{figure*} 
\centering
\includegraphics[width=0.9\textwidth]{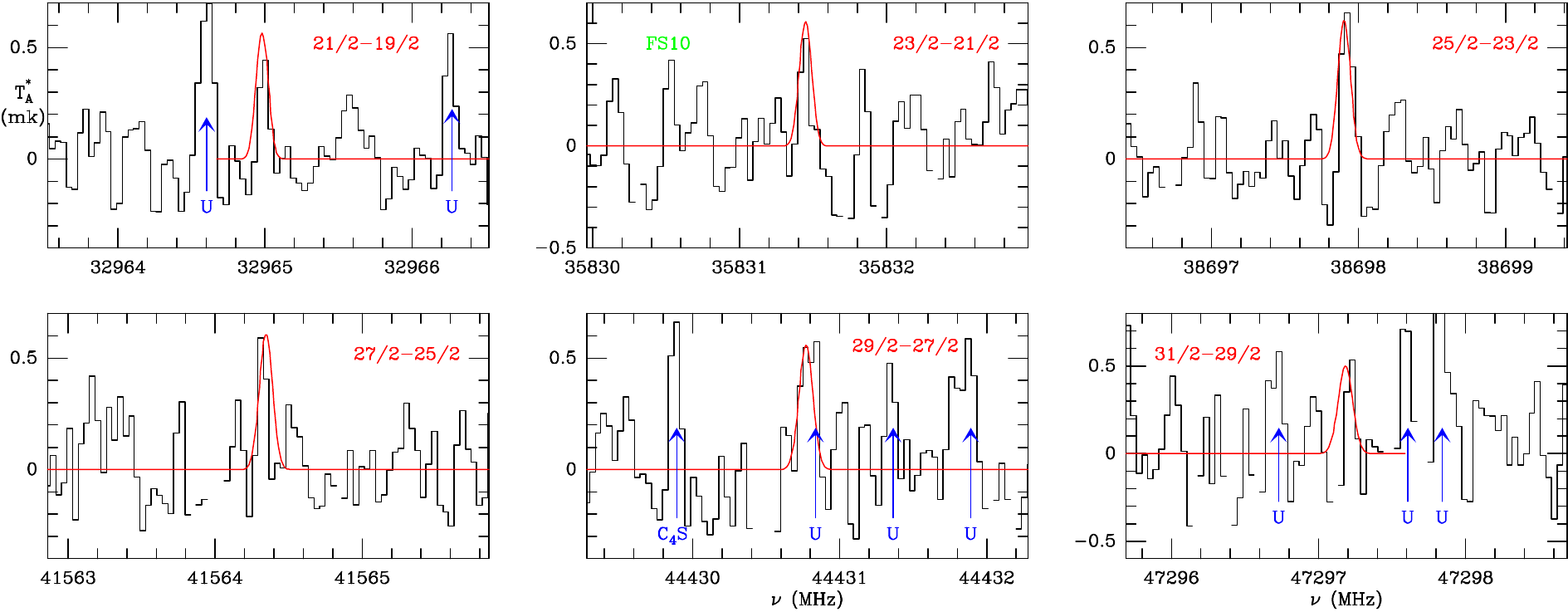}
\caption{Selected transitions of HCCCCS in TMC-1.
The abscissa corresponds to the rest frequency of the lines. Frequencies and intensities for all observed lines
are given in Table \ref{obs_line_parameters}.
The ordinate is the antenna temperature, corrected for atmospheric and telescope losses, in millikelvin.
The quantum numbers of each transition are indicated
in the corresponding panel. The red line shows the computed synthetic spectra for this species for T$_{\mathrm{rot}}$ = 10 K and
a column density of (9.5$\pm$0.8)$\times$\diez. Blanked channels correspond to negative features 
produced when folding the frequency-switched data. Green labels indicate the transitions for which only one of the frequency-switching data have been used (FS10
corresponds to a throw of 10MHz).
}
\label{HCCCCS}
\end{figure*}

\subsection{Discovery of HCCCCS} 
Butadiynethionyl (HCCCCS) is a planar molecule with a $^2\Pi$ ground state. Line frequencies and 
the value of the dipole moment used are from the CDMS catalogue. The laboratory study of HCCCCS 
was carried out by \citet{Hirahara1994}, who determined the value of B = 1434.298 MHz. The value of 
the dipole moment is $\mu$=1.45 D. We used these laboratory data to 
implement HCCCCS in the MADEX code and predict its rotational lines. 
We detected a total of six transitions, whose intensities ranged from 0.49 
to 0.68\,mK, as shown in Fig. \ref{HCCCCS}. The derived line parameters are given in Appendix 
\ref{line_parameters}.
We obtained a rotational temperature of 10$\pm$1 K and a column density of 
$N$=(9.5$\pm$0.8)$\times$10$^{10}$ cm$^{-2}$. 
Smaller values of T$_{rot}$ underestimate the intensity of the transitions at the
highest frequencies.
These derived parameters were used
to obtain the synthetic spectrum shown in Fig. \ref{HCCCCS}.
Within the family HC$_n$S, the species HCS and HCCS have been detected with column densities of 
5.5$\pm$0.5$\times$\doce$~$  and 6.8$\pm$0.6$\times$\once, respectively \citep{Cernicharo2021e}. 
Therefore, the abundance ratios HCS/HCCCCS and HCCS/HCCCCS are 58 and 7.2, respectively.  
Propadienthionyl, HC$_3$S, has not been detected yet in the ISM. From the
QUIJOTE line survey, we derived a 3$\sigma$ upper limit to its column density of 5$\times$\diez.

\section{Chemical model} \label{discussion}

\begin{figure} 
\centering
\includegraphics[width=\columnwidth]{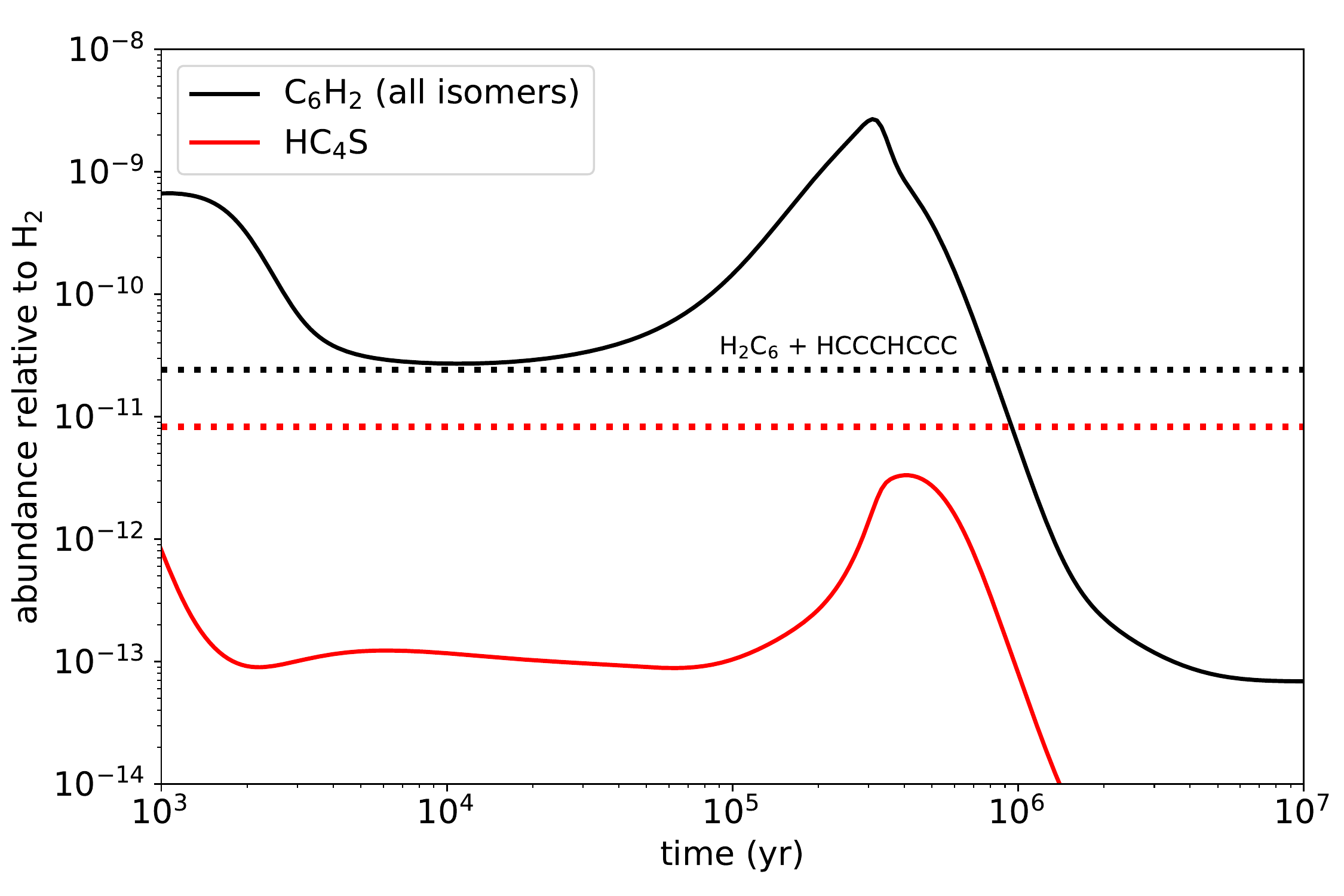}
\caption{Calculated fractional abundances of C$_6$H$_2$ (including all possible isomers) and HC$_4$S as a function of time. Horizontal dotted lines correspond to observed values in \mbox{TMC-1}; the dotted black lines account for the sum of the abundances of the two C$_6$H$_2$ isomers detected in TMC-1: H$_2$C$_6$ and HCCCHCCC. The most stable isomer, HC$_6$H, is probably the most abundant one, although it cannot be probed since it is non-polar.}
\label{fig:abun}
\end{figure}

We carried out chemical model calculations to shed light on the possible synthetic routes to HCCCHCCC and HC$_4$S. The model parameters and chemical network used are similar to those employed by \cite{Cernicharo2021e} to model the chemistry of sulphur-bearing molecules. Modifications to the chemical network are described below.

We first focused on the hydrocarbon HCCCHCCC. This species is an isomer of the C$_6$H$_2$ family. The most stable isomer, HC$_6$H, is probably the most abundant one, although it cannot be detected because it is non-polar, while the cumulene isomer H$_2$C$_6$ is detected in \mbox{TMC-1} with an abundance somewhat lower than that of HCCCHCCC. Due to the lack of information on the chemical kinetics of the different C$_6$H$_2$ isomers, the chemical networks do not differentiate between them. In the chemical model, the most important reactions of C$_6$H$_2$ formation are: C$_6$H$^-$ + H, the dissociative recombination of C$_6$H$_3^+$, and the neutral-neutral reaction C$_4$H + C$_2$H$_2$. The first and last reactions have been measured and determined to be rapid \citep{Eichelberger2007,Berteloite2010}, although information on the branching ratios of the different C$_6$H$_2$ isomers is not available. The dissociative reaction has not been measured, so the kinetic coefficient and the fragmentation channels used are an estimation determined by \citet{Herbst1989}. Given that the calculated abundance of all isomers of C$_6$H$_2$ is well above the sum of the observed abundances of H$_2$C$_6$ and HCCCHCCC (see Fig.~\ref{fig:abun}) and that the undetected isomer HC$_6$H is probably the most abundant one, it is likely that H$_2$C$_6$ and HCCCHCCC are formed by any of the aforementioned reactions, with branching ratios lower than the one that forms HC$_6$H. Information on the product distribution for these reactions would allow light to be shed on the chemistry of C$_6$H$_2$ isomers.

The sulphur-bearing species HC$_4$S is not included in chemical networks (e.g. \citealt{Vidal2017}). We thus included it assuming standard destruction routes (through reactions with neutral atoms and cations) and formation through C + H$_2$C$_3$S (in analogy to the reaction C + H$_2$CS, which is the main route to HCCS; \citealt{Cernicharo2021e}) and by the dissociative recombination of H$_2$C$_4$S$^+$, where this cation is in turn assumed to form through the reactions S + C$_4$H$_3^+$ and S$^+$ + C$_4$H$_3$. The calculated peak abundance of HC$_4$S is in reasonable agreement with the value observed in \mbox{TMC-1} (see Fig.~\ref{fig:abun}), which supports the idea that this species is formed through neutral and ionic routes similar to those that form other S-bearing carbon chains \citep{Vidal2017,Cernicharo2021e}.

\section{Conclusions}
We report on the detection of HCCCHCCC and HCCCCS towards the cold dark cloud TMC-1. For the former, we derived a column density of (1.3$\pm$0.2)$\times$\once$~$ with T$_{\mathrm{rot}}$ = 6$\pm$1 K. HCCCHCCC is thus more abundant than its cumulene isomer, which allows the chemistry of the C$_6$H$_2$ family to be constrained. For the sulphur-bearing molecule, we obtained a column density of (9.5$\pm$0.8)$\times$\diez$~$
with T$_{\mathrm{rot}}$ = 10 K. According to our chemical model, this abundance can be correctly approximated with the standard destruction and formation routes.
\begin{acknowledgements}

We thank Ministerio de Ciencia e Innovaci\'on of Spain (MICIU) for funding support through projects
PID2019-106110GB-I00, PID2019-107115GB-C21 / AEI / 10.13039/501100011033, and
PID2019-106235GB-I00. We also thank ERC for funding
through grant ERC-2013-Syg-610256-NANOCOSMOS. M.A. thanks MICIU for grant RyC-2014-16277.
\end{acknowledgements}

\normalsize

\clearpage
\onecolumn

\begin{appendix}
\section{Observed line parameters}
\label{line_parameters}

We obtained the observed line parameters for the two molecules detected in this work 
by fitting a Gaussian line profile to the observed data, using CLASS (GILDAS package). For 
this purpose, we considered a window of $\pm$15 km s$^{-1}$ around the V$_{LSR}$ (5.83 km s$^{-1}$) of the source for each transition. The results are shown in Table \ref{obs_line_parameters}. Observed lines of HCCCHCCC are shown in Figs. \ref{HCCCHCCC_k01} and \ref{HCCCHCCCk2}, and lines of HCCCCS in Fig. \ref{HCCCCS}.
Some of these parameters correspond to lines with only frequency-switching data with a throw of 8 or 10 MHz. In other cases, the lines are partially blended, but we can derive good parameters. For lines that are marginally detected, an upper limit of 3$\sigma$ is given (see Notes).

\begin{table}[h]
\caption{Observed line parameters for the species studied in  this work.} \label{obs_line_parameters}
\centering
\begin{tabular}{cccccl} 
\hline
Transition     &$\nu_{rest}$~$^a$    & $\int T_A^* dv$~$^b$&  \multicolumn{1}{c}{$\Delta$v$^c$}    & \multicolumn{1}{c}{$T_A^*$~$^d$} & Notes  \\
                   &  (MHz)              & (mK\,km\,s$^{-1}$)  &  \multicolumn{1}{c}{(km\,s$^{-1}$)}    & \multicolumn{1}{c}{(mK)}       &  \\
\hline
\multicolumn{6}{c}{\bf HCCCHCCC$^1$} \\
10$_{1,10}$-9$_{1,9}$   & 31608.167$\pm$0.010 & 1.08$\pm$0.17 & 1.20$\pm$ 0.26 & 0.85$\pm$0.13 \\
10$_{0,10}$-9$_{0,9}$   & 32160.402$\pm$0.010 & 0.81$\pm$0.11 & 1.00$\pm$ 0.16 & 0.76$\pm$0.11 \\
10$_{2,9}$-9$_{2,8}$    & 32251.428$\pm$0.020 & 0.25$\pm$0.11 & 0.74$\pm$ 0.27 & 0.31$\pm$0.13 & B \\
10$_{2,8}$-9$_{2,7}$    & 32351.748$\pm$0.010 & 0.44$\pm$0.08 & 0.64$\pm$ 0.13 & 0.64$\pm$0.11 \\
10$_{1,9}$-9$_{1,8}$    & 32867.458$\pm$0.010 & 0.92$\pm$0.11 & 0.77$\pm$ 0.11 & 1.12$\pm$0.15 & B\\
11$_{1,11}$-10$_{1,10}$ & 34763.081$\pm$0.010 & 0.70$\pm$0.14 & 0.87$\pm$ 0.20 & 0.75$\pm$0.15 \\
11$_{0,11}$-10$_{0,10}$ & 35352.883$\pm$0.010 & 0.95$\pm$0.14 & 1.13$\pm$ 0.18 & 0.79$\pm$0.12 & B \\
11$_{2,10}$-10$_{2,9}$  & 35471.993$\pm$0.010 & 0.57$\pm$0.09 & 0.64$\pm$ 0.10 & 0.83$\pm$0.11 \\
11$_{2,9}$-10$_{2,8}$   & 35605.414$\pm$0.010 & 0.76$\pm$0.10 & 1.07$\pm$ 0.16 & 0.67$\pm$0.13 \\
11$_{1,10}$-10$_{1,9}$  & 36147.207$\pm$0.010 & 0.22$\pm$0.07 & 0.53$\pm$ 0.15 & 0.38$\pm$0.12 \\
12$_{1,12}$-11$_{1,11}$ & 37916.331$\pm$0.010 & 0.42$\pm$0.11 & 0.63$\pm$ 0.18 & 0.63$\pm$0.20 & C \\
12$_{0,12}$-11$_{0,11}$ & 38538.774$\pm$0.015 & 0.60$\pm$0.14 & 1.15$\pm$ 0.36 & 0.49$\pm$0.13 & B\\
12$_{2,11}$-11$_{2,10}$ & 38691.356$\pm$0.010 & 0.41$\pm$0.09 & 0.79$\pm$ 0.19 & 0.49$\pm$0.13 \\
12$_{2,10}$-11$_{2,9}$  & 38864.149$\pm$0.010 & 0.41$\pm$0.13 & 0.49$\pm$ 0.13 & 0.78$\pm$0.20 & B\\
12$_{1,11}$-11$_{1,10}$ & 39424.853$\pm$0.015 & 0.40$\pm$0.08 & 0.67$\pm$ 0.14 & 0.56$\pm$0.14 \\
13$_{1,13}$-12$_{1,12}$ & 41067.952$\pm$0.010 & 0.54$\pm$0.11 & 0.75$\pm$ 0.16 & 0.67$\pm$0.17 \\
13$_{0,13}$-12$_{0,12}$ & 41717.861$\pm$0.010 & 0.49$\pm$0.10 & 0.47$\pm$ 0.12 & 0.98$\pm$0.14 \\
13$_{2,12}$-12$_{2,11}$ & 41909.344$\pm$0.015 & 0.29$\pm$0.10 & 0.49$\pm$ 0.25 & 0.56$\pm$0.15 \\
13$_{2,11}$-12$_{2,10}$ & 42128.143$\pm$0.010 & 0.26$\pm$0.06 & 0.34$\pm$ 0.15 & 0.70$\pm$0.21 & C \\
13$_{1,12}$-12$_{1,11}$ & 42700.173$\pm$0.010 & 0.26$\pm$0.08 & 0.36$\pm$ 0.27 & 0.69$\pm$0.17 \\
14$_{1,14}$-13$_{1,13}$ & 44217.650$\pm$0.010 & 0.26$\pm$0.09 & 0.46$\pm$ 0.16 & 0.53$\pm$0.17 \\
14$_{0,14}$-13$_{0,13}$ & 44889.725$\pm$0.015 & 0.28$\pm$0.11 & 0.51$\pm$ 0.21 & 0.51$\pm$0.17 \\
14$_{2,13}$-13$_{2,12}$ & 45125.808$\pm$0.020 & 0.21$\pm$0.10 & 0.39$\pm$ 0.21 & 0.50$\pm$0.20 \\
14$_{2,12}$-13$_{2,11}$ & 45397.603$\pm$0.080 &               &                &               & A\\
14$_{1,13}$-13$_{1,12}$ & 45972.928$\pm$0.015 & 0.57$\pm$0.18 & 0.64$\pm$ 0.19 & 0.83$\pm$0.33 & C \\
15$_{1,15}$-14$_{1,14}$ & 47365.640$\pm$0.020 & 0.15$\pm$0.10 & 0.37$\pm$ 0.19 & 0.38$\pm$0.24 \\
15$_{0,15}$-14$_{0,14}$ & 48054.091$\pm$0.020 & 0.39$\pm$0.15 & 0.57$\pm$ 0.23 & 0.64$\pm$0.24 \\
15$_{2,14}$-14$_{2,13}$ & 48340.579$\pm$0.010 & 0.12$\pm$0.06 & 0.26$\pm$ 0.09 & 0.45$\pm$0.23 \\
15$_{1,14}$-14$_{1,13}$ & 49242.813$\pm$0.010 &               &                &   $\leq$0.57 & D \\
\hline
\multicolumn{6}{c}{\bf HCCCCS$^2$} \\
21/2-19/2 &  32964.992$\pm$  0.010 &     0.32$\pm$0.11 &   0.61$\pm$ 0.17 &     0.49$\pm$0.14 \\
23/2-21/2 &  35831.439$\pm$  0.010 &     0.27$\pm$0.08 &   0.44$\pm$ 0.20 &     0.57$\pm$0.16  & C \\
25/2-23/2 &  38697.923$\pm$  0.010 &     0.48$\pm$0.10 &   0.64$\pm$ 0.12 &     0.70$\pm$0.15 \\
27/2-25/2 &  41564.324$\pm$  0.010 &     0.27$\pm$0.05 &   0.30$\pm$ 0.12 &     0.83$\pm$0.16 \\
29/2-27/2 &  44430.777$\pm$  0.010 &     0.44$\pm$0.09 &   0.68$\pm$ 0.16 &     0.61$\pm$0.16 \\
31/2-29/2 &  47297.214$\pm$  0.010 &     0.21$\pm$0.05 &   0.25$\pm$ 0.42 &     0.78$\pm$0.22 \\
\hline
\end{tabular}
\tablefoot{\\
\tablefoottext{1}{Quantum numbers are $J'_{K'_{a,}K'_{c}}$ - $J_{K_{a,}K_{c}}$.}\\
\tablefoottext{2}{Quantum numbers are $J'$-$J$.} \\
\tablefoottext{a}{Observed frequency of the transition assuming a local standard of rest velocity of 5.83 km s$^{-1}$.}\\
\tablefoottext{b}{Integrated line intensity in mK\,km\,s$^{-1}$.}\\
\tablefoottext{c}{Linewidth at half intensity derived by fitting a Gaussian function to
the observed line profile (in km\,s$^{-1}$).}\\
\tablefoottext{d}{Antenna temperature in millikelvin.}\\
\tablefoottext{A}{This line is fully blended with a transition of CH$_2$CCHCCH. The frequency corresponds to that calculated with the laboratory constants of Table \ref{new_rotational_HCCCHCCC}.}\\
\tablefoottext{B}{Frequency-switching data with a throw of 8 MHz only. Negative feature present in the data with a 10 MHz throw.}\\
\tablefoottext{C}{Frequency-switching data with a throw of 10 MHz only. Negative feature present in the data with a 8 MHz throw.}\\
\tablefoottext{D}{Upper limit corresponds to 3$\sigma$.}\\
}
\end{table}

\twocolumn

\onecolumn
\section{Improved rotational constants for HCCCHCCC}
\label{new_constants}

We merged the laboratory lines of HCCCHCCC obtained by \citet{McCarthy2002} with those observed in TMC-1. This 
improves the values of the rotational and distortion constants. The results are given in Table \ref{new_rotational_HCCCHCCC}. 
The values obtained from the merged fit are very similar to those obtained from the laboratory fit, with the 
greatest difference being less than 1 kHz.

\begin{table}[h]
\caption{Improved rotational and distortion constants for HCCCHCCC.}
\label{new_rotational_HCCCHCCC}
\centering
\begin{tabular}{{|c|c|c|c|}}
\hline
Constant                 & Theoretical$^1$ & Laboratory$^2$ & Lab + TMC-1$^3$ \\
(MHz)                    &                &                 &                 \\
\hline
$A  $                    & 21399.8 & 21094.10161(25)   &  21094.10153(52)\\
$B  $                    & 1664.0  &  1676.268423(70)  &   1676.26835(81)\\
$C  $                    & 1543.9  &  1550.066623(42)  &   1550.066695(60)\\
$\Delta_J$\,$\times$10$^3$    & 0.579   &    0.6414(77)     &  0.64165(44)\\
$\Delta_{JK}$                 & -0.0662 &   -0.075808(14)   &  -0.075820(25) \\
$\delta_J$\,$\times$10$^3$    & 0.144   &    0.16067(50)    &   0.15976(52)  \\
\hline
\hline
$J_{max}$                &          &      7           &  15    \\    
$K_{a,max}$              &          &      1           &  2    \\
$N_{lines}$              &          &       24         & 49     \\
$\sigma$ (kHz)           &          &      0.5         & 1.5    \\
$\nu_{max}$ (MHz)        &          &    40330         & 48340    \\
\hline
\end{tabular}
\tablefoot{\\
        \tablefoottext{1}{Ab initio calculation from \citet{sattelmeyer2000}.}\\
        \tablefoottext{2}{Rotational and distortion constants derived from a
       fit to the laboratory data of \citet{McCarthy2002}.}\\
        \tablefoottext{3}{Rotational and distortion constants derived from a merged fit to
       the laboratory and TMC-1 frequencies.}\\
}
\end{table}
\normalsize

\end{appendix}


\begin{thebibliography}{} 
\tiny
\bibitem[Ag\'undez et al. (2021)]{Agundez2021} Ag\'undez, M., Cabezas, C., Tercero, B., et al. 2021, \aap, 647, L10 %H2CCCH
\bibitem[Berteloite et al. (2010)]{Berteloite2010} Berteloite, C., Le Picard, S., Balucani, N., et al. 2010, PCCP, 12, 3677
\bibitem[Cabezas et al. (2021)]{Cabezas2021} Cabezas, C., Tercero, B., Agúndez, M. et al. 2021, \aap, 650, L9 %H2C5
\bibitem[Cabezas et al. (2022)]{Cabezas2022} Cabezas, C., Ag\'undez, M., Marcelino, N., et al. 2022, 657, L4 %HCCS+
\bibitem[Cernicharo(1985)]{Cernicharo1985} Cernicharo, J. 1985, Internal IRAM report (Granada: IRAM)
\bibitem[Cernicharo et al. (1991a)]{Cernicharo1991} Cernicharo, J., Gottlieb, C.A., Guélin, M., et al. 1991, \apj , 368, L39
\bibitem[Cernicharo et al. (1991b)]{Cernicharo1991b} Cernicharo, J., Gottlieb, C. A., Guélin, M., et al. 1991b, \apj, 368, L43
\bibitem[Cernicharo \& Gu\'elin(1987)]{Cernicharo1987} Cernicharo, J. \& Gu\'elin, M. 1987, \aap, 176, 299
\bibitem[Cernicharo (2012)]{Cernicharo2012} Cernicharo, J., 2012, in ECLA 2011: Proc. of the European Conference on Laboratory Astrophysics,
EAS Publications Series, 2012, Ed.: C. Stehl, C. Joblin, \& L. d'Hendecourt (Cambridge: Cambridge Univ. Press),
251; \texttt{https://nanocosmos.iff.csic.es/?page$\_$id=1619}
\bibitem[Cernicharo et al. (2018)]{Cernicharo2018} Cernicharo, J., Guélin, M., Agúndez, M., et al. 2018, \aap, 618, A4
\bibitem[Cernicharo et al.(2021a)]{Cernicharo2021a} Cernicharo, J., Ag\'undez, M., Kaiser, R., et al. 2021a, \aap, 652, L9 %benzyne
\bibitem[Cernicharo et al.(2021b)]{Cernicharo2021b} Cernicharo, J., Ag\'undez, M., Cabezas, C., et al. 2021b, \aap, 647, L2  %CH2CHCCH
\bibitem[Cernicharo et al.(2021c)]{Cernicharo2021c} Cernicharo, J., Ag\'undez, M., Cabezas, C., et al.  2021c, \aap, 649, L15 %indeno
\bibitem[Cernicharo et al.(2021d)]{Cernicharo2021d} Cernicharo, J., Cabezas, C., Endo, Y., et al. 2021d, \aap, 646, L3 %HC3S+
\bibitem[Cernicharo et al.(2021e)]{Cernicharo2021e} Cernicharo, J., Cabezas, C., Ag\'undez, M., et al. 2021e, \aap, 648, L3 %NCS HCCS ...
\bibitem[Cernicharo et al.(2021f)]{Cernicharo2021f} Cernicharo, J., Cabezas, C., Endo, Y., et al. 2021f, \aap, 650, L1 %HCSCN & HCSCCH
\bibitem[Cernicharo et al.(2022)]{Cernicharo2022} Cernicharo, J., Fuentetaja, R., Ag\'undez, M., et al. 2022, \aap, 663, L9
\bibitem[Eichelberger et al.(2007)]{Eichelberger2007} Eichelberger, B., Snow, T. P., Barckholtz, C., \& Bierbaum, V. M. 2007, \apj, 667, 1283
\bibitem[Foss\'e et al. (2001)]{Fosse2001} Foss\'e, D., Cernicharo, J., Gerin, M., Cox, P. 2001, \apj, 552, 168
\bibitem[Fuentetaja et al. (2022)]{Fuentetaja2022} Fuentetaja, R., Cabezas, C., Ag\'undez, M., et al. 2022, \aap,  663, L3
\bibitem[Gu\'elin et al. (2000)]{Guelin2000} Guélin, M., Müller, S., Cernicharo, J. 2000, \aap, 363, L9
\bibitem[Herbst et al. (1989)]{Herbst1989} Herbst, E., \& C.M., Leung 1989, \apj, 69, 271
\bibitem[Hirahara et al. (1994)]{Hirahara1994} Hirahara, Y., Ohshima, Y., \& Endo, Y. 1994, J. Chem. Phys., 101, 7342
\bibitem[Kawaguchi et al. (1991)]{Kawaguchi1991} Kawaguchi, K., Kaifu, N., Ohishi, M., et al. 1991, PASJ, 43, 607
\bibitem[Langer et al. (1997)]{Langer1997} Langer, W. D., Velusamy, T., Kuiper, T. B. H., et al. 1997, \apj, 480, L63
\bibitem[McCarthy \& Thaddeus (2002)]{McCarthy2002} McCarthy. M.C., \& Thaddeus, P. 2002, \apj, 569, L55 %Laboratory data
\bibitem[McGuire et al.(2018)]{McGuire2018} McGuire, B.A., Burkhardt, A.M., Kalenskii, S., et al. 2018, Science, 359, 202
\bibitem[M\"uller et al.(2005)]{Muller2005} M\"uller, H.S.P., Schl\"oder, F., Stutzki, J., Winnewisser, G. 2005, \jmst, 742, 215
\bibitem[Pardo et al.(2001)]{Pardo2001} Pardo, J.~R., Cernicharo, J., Serabyn, E. 2001, IEEE Trans. Antennas and Propagation, 49, 12
\bibitem[Pickett et al. (1998)]{Pickett1998} Pickett, H.M., Poynter, R.~L., Cohen, E.~A., et al. 1998, J. Quant. Spectrosc. Radiat. Transfer, 60, 883
\bibitem[Saito et al. (1987)]{Saito1987} Saito, S., Kawaguchi, K., Yamamoto, S., et al. 1987, \apj, 317, L1156
\bibitem[Sattelmeyer \& Stanton (2000)]{sattelmeyer2000} Sattelmeyer, K.W., \& Stanton, J.F. 2000, J. Am. Chem. Soc., 122, 8220
\bibitem[Tercero et al.(2021)]{Tercero2021} Tercero, F., L\'opez-P\'erez, J. A., Gallego, et al. 2021, \aap, 645, A37
\bibitem[Thaddeus et al. (1985)]{Thaddeus1987} Thaddeus, P., Vrtilek, J.M., \& G\"uttlieb, C.A. 1987, \apj, 299, L63
\bibitem[Vidal et al.(2017)]{Vidal2017} Vidal, T. H. G., Loison, J.-C., Jaziri, A. Y., et al. 2017, \mnras, 469, 435
\bibitem[Yamamoto et al. (1987)]{Yamamoto1987} Yamamoto, S., Saito, S., Kawaguchi, K., et al. 1987, \apj, 317, L119



\end{thebibliography}
\end{document}